\documentstyle[aps,prb,epsfig,multicol,color]{revtex}

\begin{document}
\draft

\title{ $1/T_1$ in the D-wave superconducting state with  coexisting
antiferromagnetism}

\author{Yunkyu Bang$^{*}$, M. J. Graf, A.V. Balatsky, and J. D. Thompson}

\address {Los Alamos National Laboratory, Los Alamos,
New Mexico 87545 }

\date{\today}
\maketitle

\begin{abstract}
We consider a general D-wave superconducting gap function in the presence of
coexisting antiferromagnetic (AFM) long-rang order. We find that the D-wave
order parameter develops additional nodes without lowering the symmetry of the
state due to the interplay of the superconducting and AFM correlations. This
AFM correlated D-wave gap has a small gap structure inside the generic D-wave
gap. As a result of this additional structure it shows a quite different
response to impurities compared to a standard D-wave gap.  We calculate the
nuclear spin-lattice relaxation rate $1/T_1$ of this gap in the presence of
impurities and discuss the implications for the experiments of some AFM heavy-
fermion superconductors and in particular for CeRhIn5.
\end{abstract}



\begin{multicols}{2}

\section{Introduction}

The interplay and competition between antiferromagnetism (AFM) and
superconductivity (SC) is a long studied subject both experimentally
\cite{Stewart} and theoretically \cite{theory}.  There are now several heavy
fermion compounds showing clear experimental evidence of the microscopically
homogeneous coexistence of AFM and SC correlations
\cite{CeCu2Si2,URu2Si2,CeCu2Si2Ge,UPd2Al3}. However, the direct consequences of
the coexistence of AFM and SC has not yet been clearly resolved. Recent
$^{115}$In- NQR studies  of CeRhIn$_5$ and CeIn$_5$ also revealed that there is
a coexistence region of AFM and SC around the critical pressures of 1.6 GPa and
2.43 GPa, respectively \cite{Kitaoka}. Apart from confirming the coexistence of
AFM and SC in this region of pressure, an interesting observation is that the
nuclear spin-lattice relaxation rate $1/T_1$  displays the linear temperature
dependence $1/T_1\propto$ T at low temperatures (up to 0.3 T$_{sc}$ for
CeRhIn$_5$), followed by the T$^3$ behavior below the superconducting
transition temperature T$_{sc}$. The latter behavior is an indication of the
lines of nodes of the anisotropic SC state.

The standard explanation of the T-linear temperature dependence of $1/T_1$ at
low temperatures is the impurity effect in an unconventional SC with lines of
nodes. However, this  impurity explanation  has  difficulties in the case of
CeRhIn$_5$; the low temperature  T-linear $1/T_1$ is only observed in the AFM
and SC coexistence region at 1.6 GPa, and once  the AFM order disappears by
increasing pressure to 2.1 GPa,   $1/T_1$ exhibits a T$^3$ behavior for the
entire measured temperatures below T$_{sc}$ \cite{Mito}. Similar pressure
dependence of $1/T_1$ has also been observed in Ge-substituted CeCu$_2$Si$_2$
\cite{CeCu2Si2Ge}.  Another AFM heavy fermion superconductor URu$_2$Si$_2$
\cite{URu2Si2} also shows the low temperature T-linear behavior in $1/T_1$. In
the case of UPd$_2$Al$_3$, the situation is somewhat different; a more recent
NMR experiment \cite{UPd2Al3} indicates that the T-linear behavior observed in
earlier experiments disappears with purer samples. Putting together these
observations we think that the interplay of the AFM and SC, together with some
sort of impurity effects, should hold the key to explain the $1/T_1 \propto$ T
behavior at low temperatures in these AFM heavy fermion superconductors. In
particular, we think this is the case for CeRhIn$_5$ under pressure, where at
low temperatures  $1/T_1 \propto$ T is observed in the coexistence region
(P=1.6 GPa), but is replaced by $1/T_1 \propto$ T$^3$ once the AFM is
suppressed (P=2.1GPa).

There is a substantial body of studies  on the interplay/coexistence between
D-wave SC and AFM \cite{theory}. The general conclusion of these studies is
that, depending on the parameters in the model, these two phases either exclude
each other or can coexist together.
Here, however, we are motivated by the  above-mentioned magnetic heavy-fermion
(HF) superconducting materials, where the Ne\'el temperature  $T_N$ usually
sets in first at a higher temperature and then at a lower temperature the SC
transition occurs on top of the AFM ordering. Henceforth, we assume the
coexistence of D-wave SC and AFM and confine our interest on how a general
D-wave SC order parameter (OP) should be modified due to the presence of the
AFM long range order.
We found an important modification of the D-wave gap due to the long range AFM
order; the D-wave SC state develops additional nodes in the gap function that
are a direct consequence of the crossing between the Fermi surface (FS) and the
magnetic Brillouin zone (BZ), see Fig.1. At these discrete crossing points the
superconducting order parameter is forced to vanish by the AFM coherence
factors \cite{Lev}.
These additional nodes (we call them {\em AFM nodal points}) lead to a density
of states (DOS), which is almost gapless down to  very low energy before it
develops a steeply decreasing DOS, see Fig.2. This shape of the DOS is
intrinsically vulnerable to a small amount of impurities, which fill rapidly
the small gap and  produce a large finite DOS.

From this AFM+D-wave gap solution we calculate the nuclear spin-lattice
relaxation rate $1/T_1$ in the presence of  impurities. We find that the
response of this AFM+D-wave gap to impurities is quite different from the one
of the standard D-wave gap. The main findings are: (1) the AFM+D-wave gap
produces more quickly a finite DOS at $\omega=0$ than the standard D-wave gap,
both for Born and for unitary impurity scattering, (2) for unitary scattering,
the calculated $1/T_1$, both for the AFM+D-wave gap and for the standard D-wave
gap, produces the  T-linear behavior at very low temperatures. The temperature
regions of T-linear behavior are about the same  for both gaps with the same
amount of impurity concentration despite the different low energy behaviors of
the DOS. However, the magnitude of $1/T_1$ in the T-linear region is about 4
times smaller for the standard D-wave gap than for the AFM+D-wave gap, (3) for
Born limit scattering,  the $1/T_1$ of the AFM+D-wave gap can still produce the
T-linear behavior in a substantial region of low temperatures but with a much
higher impurity concentration than with unitary impurities.  In order to
achieve the same T-linear behavior as for unitary scattering, about 5 times the
amount of impurity concentration is needed.
On the other hand, the standard D-wave gap hardly shows any T-linear behavior
with the same amount of impurities, but displays almost T$^3$ power law for the
entire temperature range below T$_{sc}$ with  a small deviation due to
impurities.

Considering  the experimental observation of the T-linear behavior in $1/T_1$
in some of the AFM heavy fermion superconductors,  we propose that the
AFM+D-wave gap structure together with Born limit impurities can be the reason
for the T-linear behavior in $1/T_1$ in these compounds.  When the AFM long
range order disappears by tuning pressure  in CeRhIn$_5$, as well as in
CeCu$_2$(Si$_{0.98}$Ge$_{0.02}$)$_2$, the SC gap gradually returns to a
standard D-wave gap and then a small amount of Born scatterers is not enough to
produce the T-linear $1/T_1$ \cite{CeCu2Si2Ge,Kitaoka}. One remaining question
of this scenario is the possible source of the Born limit scatterers. We think
that the most probable source of scatterers  in the  Born limit is  strain in
the lattice induced with pressure and/or AFM domain walls  in CeRhIn$_5$ and
through chemical substitution in CeCu$_2$(Si$_{0.98}$Ge$_{0.02}$)$_2$
\cite{CeCu2Si2Ge}.

\section{Formalism}

The Hamiltonian describing the coexistence of AFM and SC pairing interactions
is written as

\begin{eqnarray}
H &=& \sum_{k,\sigma} [\epsilon_k c\dag _{k,\sigma} c_{k,\sigma} + \sigma
\Delta_{M} c\dag _{k+Q,\sigma} c_{k,\sigma}] \nonumber \\ & & +\sum_{k,k'}
V(k,k') c\dag _{k,\sigma} c\dag _{-k,\sigma'} c _{-k',\sigma'} c _{k',\sigma},
\end{eqnarray}
\noindent where we  replaced the AFM pairing interaction by its mean-field AFM
OP $\Delta_{M}$. In the presence of  the AFM order, the conduction band is
split into two AFM quasiparticle bands:

\begin{eqnarray}
\alpha_{(1),\sigma,k} &=& u_k c_{\sigma,k} -\sigma v_k c_{\sigma,k+Q},
\nonumber
\\
\alpha_{(2),\sigma,k} &=& \sigma v_k c_{\sigma,k} + u_k c_{\sigma,k+Q},
\end{eqnarray}

\noindent with the AFM Bogoliubov coefficients, $u_k ^2= 1/2 +
\frac{\delta_k}{2 \sqrt{\delta_k ^2 + \Delta_M ^2}}$, and $v_k ^2= 1/2 -
\frac{\delta_k}{2 \sqrt{\delta_k ^2 + \Delta_M ^2}}$. The quasiparticle
energy
and the convenient parameters are defined as

\begin{eqnarray}
E_{(1),(2)} &=& \zeta_k \pm \sqrt{\delta_k ^2 + \Delta_M ^2}, \\ \zeta_k & =&
\frac{1}{2} [\epsilon_k +\epsilon_{k+Q}], \\ \delta_k &=& \frac{1}{2}
[\epsilon_k -\epsilon_{k+Q}].
\end{eqnarray}

Then the original SC paring interaction is rewritten in terms of  the AFM
quasiparticle wave function,

\begin{eqnarray}
H_{int} &=& \sum _{k,k'} V(k,k') c\dag _{\uparrow,k} c\dag _{\downarrow,-k}
c_{\downarrow,-k'} c_{\uparrow,k'} \nonumber \\ & =&  \sum _{k,k'} V(k,k') (u_k
^2 -v_k ^2)  \alpha \dag _{(2) \uparrow,k} \alpha \dag _{(2) \downarrow,-k}
\nonumber
\\ && \alpha_{(2) \downarrow,-k'} \alpha_{(2) \uparrow,k'}
(u_{k'} ^2 -v_{k'} ^2) + ...,
\end{eqnarray}

\noindent where the  ellipsis stands for  terms like $ \alpha_1 \dag \alpha_1
\dag _1 \alpha_1 \alpha _1$ and all the mixed terms of $ \alpha_1$ and $ \alpha
_2$. In this approximation we certainly keep the pairing term in the lower AFM
quasiparticle band assuming that the Fermi level crosses the lower band and
that the AFM gap $\Delta_M$ is much larger than the SC gap $\Delta_{sc}$. This
approximation should apply for the case $T_N \gg T_{sc}$. Then the singlet gap
equation is written as

\begin{eqnarray}
\Delta_{\alpha}(k) &=& (u_k ^2 -v_k ^2) \sum_{k'} V(k,k') <\alpha_{(2)
\uparrow,k'} \alpha_{(2) \downarrow,-k'}> \nonumber \\ && \times (u_{k'}
^2-v_{k'} ^2).
\end{eqnarray}

This gap equation was obtained by Buzdin and Bulaevskii \cite{Lev} for s-wave
SC. The important point is that the AFM coherence factor $(u_k ^2 -v_k ^2)$
modifies the otherwise standard D-wave gap equation where SC is mediated by the
pairing potential $V(k,k')$.  At those points where $(u_k ^2 -v_k ^2)=0$ is
satisfied, the gap function $\Delta_{\alpha}(k) $ should vanish in addition to
the D-wave nodal points. Now for simplicity of numerical calculations, we
assume a circular FS in two dimensions and the gap equation is generalized to
account for impurity effects \cite{Bang}.
The effect of the impurity scattering is included with T-matrix
approximation\cite{Hirschfeld}. For the particle-hole symmetric case $T_3=0$,
and for D-wave OP with isotropic scattering $T_1=0$ (also without loss of
generality we can choose $T_2=0$ by U(1) symmetry). Then we need to calculate
only $T_0(\omega)$. The impurity selfenergy is given by $\Sigma_{0}=\Gamma
T_{0}$, where $\Gamma=n_i/\pi N_{0}$, $N_0$ is the normal DOS at the Fermi
energy, $n_i$ is the impurity concentration;  $T_0 (\omega_n) =\frac{g_0
(\omega_n)}{[c^2-g_0 ^2 (\omega_n)]}$, where $g_0 (\omega_n) = \frac{1}{\pi
N_0}  \sum_k \frac{i \tilde{\omega}_n}{\tilde{\omega}_n^2 + \epsilon_k^2
+\Delta^2(k)}$, $\tilde{\omega}_n=\omega_n+\Sigma_0$ ($\omega_n=\pi$ T
$(2n+1)$), and the scattering strength parameter $c$ is related to the s-wave
phase shift $\delta$  by  $c=\cot(\delta)$.
With this $T_0$ the following gap equation is solved self-consistently.
\begin{eqnarray}
\Delta(\phi) &=& - N(0) \cdot g(\phi) \int \frac{d \phi^{'}}{2 \pi}
V(\phi-\phi^{'})  \nonumber
\\ & & \times  T \sum_{\omega_n} \int^{\omega_D}_{-\omega_D} d \epsilon
\frac{
\Delta(\phi^{'})}{\tilde{\omega}_n^2 + E_{(2)} ^2 (\epsilon)
+\Delta^2(\phi^{'})} \cdot g(\phi^{'}),
\end{eqnarray}

\noindent where $g(\phi)$ is the angular parametrization of $(u_k ^2 -v_k
^2)$.
In contrast to previous works on the D-wave SC, where the form of the D-wave SC
gap has a fixed functional form such as $\Delta(\vec{k})= \Delta_0 (\cos k_x
-\cos k_y)$ or $\cos(2 \phi)$, we allow for a most general D-wave OP of  $D_2$
symmetry; namely $\Delta(n\pi/4)=0 (n=1,3,5,7) $, $\Delta(\phi)=\Delta(\phi\pm
\pi)$, and $\Delta(\phi)= -\Delta(\phi\pm \pi/2)$.
Therefore the gap equation can produce the most general D-wave symmetry gap
solution for a given pairing potential and now with an additional constraint
from the AFM correlation. The pairing potential $V(\phi-\phi^{'})$  induces a
D-wave gap. Although its microscopic origin is not an issue in this paper, we
believe it originates from  AFM fluctuations. The static limit of AFM
fluctuations $\chi({\bf q},\omega=0) \sim \frac{1}{(q-Q)^2+\xi^{-2}}$ is
parameterized as \cite{Bang,pairing potential}

\begin{equation}
V(\phi-\phi^{'})=V_d(b) \frac{b^2}{(\phi-\phi^{'} \pm \pi/2)^2+b^2}
\end{equation}

\noindent where the parameter $b$ is proportional to $\xi^{-1}$, normalized in
the circular FS $(\xi \sim a \pi/ b$; $a$ is the lattice parameter). For all
calculations in this paper, we choose $b=0.5$ which is not a sensitive
parameter for our results.

With the gap function $\Delta(\phi)$ and  $T_0 (\omega)$ obtained from Eq.(8)
($T_0 (\omega)$ is analytically continued from $T_0 (\omega_n)$ by Pad\'e
approximant method), we calculate the $1/T_1$ nuclear spin-lattice relaxation
rate  \cite{Hirschfeld,Choi}

\begin{eqnarray}
\frac{1}{T_1 T} &\sim&  \int_0 ^{\infty} \frac{\partial f_{F}
(\omega)}{\partial \omega}   [ (\langle Re
\frac{\tilde{\omega}}{\sqrt{\tilde{\omega}^2-\Delta^2(\phi)}} \rangle_{\phi})^2
\nonumber \\ && + ( \langle Re
\frac{\Delta(\phi)}{\sqrt{\tilde{\omega}^2-\Delta^2(\phi)}} \rangle_{\phi})^2
],
\end{eqnarray}

\noindent where $\tilde{\omega}=\omega+\Sigma_0(\omega)$ and
$\langle...\rangle_{\phi}$ means the angular average over the FS. The first
term in the bracket of Eq. (10) is  $N^2 (\omega)$. The second term vanishes in
our calculations because of the symmetry of the OP.  To calculate $1/T_1 T$
using Eq. (10), we need the full temperature dependent gap function
$\Delta(\phi,T)$ and T$_{sc}$. Our gap equation Eq. (8) is basically the BCS
gap equation, therefore it gives  the BCS temperature behavior for
$\Delta(\phi,T)$ and $\Delta_0/ T_{sc}=2.14$ for the standard D-wave SC. In
order to account for strong coupling effects and a stronger anisotropy of our
general D-wave gap solution, we use the phenomenological formula,
$\Delta(\phi,T)=\Delta(\phi,T=0)~ \Xi(T)$ with $\Xi(T)=\tanh (\beta
\sqrt{T_{sc}/T-1})$, and parameters $\beta$ and $\Delta_0/ T_{sc}$.   Then we
only need to calculate $\Delta(\phi,0)$ at zero temperature. The temperature
dependence of $\Sigma_0(\omega,T)$ ($=\Gamma~ T_0 (\omega,T)$) is similarly
extrapolated: $T_0(\omega,T)=T_0(\omega,T=0)~ \Xi(T) + T_{normal}(1-\Xi(T))$,
where $T_{normal}=\Gamma/(c^2+1)$ is the normal state $T_0$. In our numerical
calculation $\beta=1.74$ is fixed because this parameter is not sensitive for
the final results. But the ratio $\Delta_0/ T_{sc}$ is an important parameter
to simulate
 strong coupling effects;  the larger this ratio is the more the strong
coupling effect is accounted for.

\section{Results and Discussions}

Now let us discuss the numerical results. In Fig. (1) the schematic FS and the
magnetic BZ are shown together with the two-dimensional original BZ. The points
where the magnetic BZ and the original FS cross and therefore the condition
$(u_k ^2 -v_k ^2)=0$ is satisfied, are marked as {\em AFM nodal points}. The
gap function vanishes at those points; in real space  this means that the
singlet SC pairing is prohibited along the (1,1) direction due to  AFM spin
correlations \cite{Lev}. A schematic D-wave gap solution is drawn accordingly.

In Fig. (2), we show the normalized DOS $N(\omega)/N_0$ for two exemplary cases
of the AFM nodal points ($\eta=0.2$ and $0.4$); the distance of the AFM nodal
point from the D-wave nodal point is parameterized by
$\eta=|\phi_{AFM}-\phi_{D}|/\phi_{D}$, where $\phi_{D}=\pi/4$ is the D-wave
nodal point and $\phi_{AFM}$ is the AFM nodal point. The inset shows the
corresponding  gap solutions. We found empirically that $\Delta (\phi) \sim [
\cos(2 \phi) + 1.5 ~\eta \cos(6 \phi)$] is a very good approximation to fit the
AFM+D-wave gap solutions.
In real compounds, the distance of the AFM nodal points from the D-wave nodal
points is determined by the shape of the original FS  (before the AFM long
range order is included). The key feature is that when the AFM nodal points are
not very far from the D-wave nodal points, the DOS already develops the gapless
feature (without impurities), except at very low energies. This very low energy
region has a shallow gap (the steeply rising DOS region), which is
intrinsically vulnerable to disorder.

Fig.3 (a) shows the DOS and $1/T_1$ calculated for the AFM+D-wave gap with
unitary impurities ($c=0$). The inset shows the normalized DOS of the
AFM+D-wave gap solution with varying  concentrations of the unitary scatterers.
We choose $\eta=0.3$ (see Fig.  2) for illustration of a typical AFM+D-wave
gap. As explained  above, the low energy region of the DOS is quickly filled
with a small amount of impurities. With this choice of parameters the impurity
scattering rate $\Gamma/\Delta_0=0.064$ is enough to completely fill the low
energy gap region and  $N(\omega=0)$ reaches more than 50 $\%$ of the
normal-state DOS $N_0$.
The main panel shows the nuclear spin-lattice relaxation rate $1/T_1$ in a
 log-log scale for the corresponding DOS in the inset. In Fig. 3 and Fig. 4  all
$1/T_1$ results are normalized to $1/T_1$=10 at T=T$_{sc}$ for easy comparison.
For the temperature dependence of the gap,  we choose the parameters
$\beta=1.74$ and $\Delta_0/ T_{sc}=3$  for the AFM+D-wave gap to account for
the strong coupling effects of superconductivity as explained before. As
expected from the DOS results, due to the impurity induced residual states,
$1/T_1$ displays the linear-T dependence at low temperatures and the region of
T-linear behavior increases with  impurity concentration. For
$\Gamma/\Delta_0=0.064$, this T-linear region extends up to $\sim 0.35$
T$_{sc}$. In the high temperature region, first of all,  near T$_{sc}$ the
coherence peak is almost invisible because of the sign-changing gap function
(the second term in Eq. (10) vanishes). Secondly, it shows the typical T$^3$
behavior below T$_{sc}$ due to the lines of nodes in the gap until it goes
through a gradual crossover region and finally to the T-linear region. The
comparison with the experimental data  of CeRhIn$_5$ at 1.6 GPa \cite{Kitaoka}
(green diamond symbols; the experimental data are normalized  in the same
fashion as the numerical results) is quite good with the theoretical
calculation with $\Gamma/\Delta_0=0.064$. On the other hand, there is no way to
fit the 2.1GPa data (magenta diamonds) with any amount of impurities.

Fig.3 (b) shows the same calculations as in Fig.3 (a) for a standard harmonic
D-wave ($\Delta(\phi)=\Delta_0 \cos(2\phi)$) for comparison. The parameters
$\beta=1.74$ and $\Delta_0/ T_{sc}=2.5$ are used for the harmonic D-wave case.
Compared to the AFM+D-wave case, we  notice the main difference in the DOS at
low energies shown in the inset. The resonance feature of unitary scattering in
the residual DOS $N(\omega \rightarrow 0)$ is seen more clearly but the value
of $N(\omega=0)$ is smaller than for the AFM+D-wave case (see Fig.3 (a)) with
the same impurity  scattering rate.
As in the AFM+D-wave case,  $1/T_1$  produces the T-linear behavior at low
temperatures with impurities  and the T$^3$ dependence at higher temperatures
below T$_{sc}$.
However, there are also some noticeable differences compared to the AFM+D-wave
case. First, the T-linear region and the T$^3$ region are more clearly
separated (a smaller crossover region compared to the AFM+D-wave gap case).
This difference  reflects  the different shapes of the DOS for each case.
Second, another more important difference is that the magnitude of the low
temperature $1/T_1$ is about four times smaller for the harmonic D-wave than
the one of the AFM+D-wave case with the same impurity concentration. Therefore,
the comparison with the data of CeRhIn$_5$ at 1.6 GPa \cite{Kitaoka} (green
diamonds) shows a large deviation at low temperatures although the T-linear
power law is reproduced. It is possible to fit the 1.6 GPa data with a much
larger impurity scattering rate of $\Gamma/\Delta_0 \sim 0.16$ (not shown in
Fig.3 (b)  but which can be extrapolated from the shown results). On the other
hand, in order to fit the 2.1 GPa data (magenta diamonds), $\Gamma/\Delta_0$
needs to be 0.016 or smaller (see Fig.3 (b)). Thus, we would need an
explanation for why the impurity scattering rate is suddenly reduced by more
than one order of magnitude when changing pressure from 1.6 GPa to 2.1 GPa.

Summing up the results of our calculations for unitary impurity scattering, we
found that both gap functions with unitary impurities have difficulties to
explain consistently the experiments at 1.6 GPa and 2.1 GPa.

Now we consider the effects of Born  impurity scattering. Fig.4 (a) shows the
DOS and $1/T_1$ of the AFM+D-wave gap with Born limit impurities ($c=1$). The
inset of Fig.4 (a) shows the DOS of the AFM+D-wave gap ($\eta=0.3$) with
different impurity concentrations.  Compared to the unitary impurity case
(Fig.3.(a)), the shape of the DOS looks almost the same except for the fact
that the amount of impurity concentration needs to be about five times larger
to achieve a similar residual DOS.  As a result, $1/T_1$  looks similar as in
the unitary impurity case (Fig.3 (a)) with five times larger impurity
concentrations. We can fit the 1.6 GPa data of CeRhIn$_5$  with the Born
impurity scattering rate $\Gamma /\Delta_0=0.32$.

However, the results for the harmonic D-wave gap with Born impurities are quite
different. Fig.4 (b) shows the DOS and $1/T_1$ for the harmonic D-wave case
with Born limit scatterers. With increasing impurity  scattering rate up to
$\Gamma /\Delta_0=0.32$, the zero frequency residual DOS $N(0)$ is  less than
20$\%$ of $N_0$. As a result $1/T_1$ never displays any noticeable T-linear
behavior up to $\Gamma /\Delta_0=0.32$. On the other hand, the comparison with
the 2.1 GPa data of CeRhIn$_5$ \cite{Kitaoka,Mito} shows that the experimental
data can be fitted nicely with  $\Gamma /\Delta_0=0.16$.

A consistent explanation of the experiments of CeRhIn$_5$ with our calculations
is the following. We first estimate $\Delta_{0,2.1 GPa} / \Delta_{0,1.6 GPa}
\sim 2$ from the experimental observation that T$_{sc;2.1 GPa} \sim 2.3$K and
T$_{sc;1.6 GPa} \sim 1$ K. Then at 1.6 GPa where the AFM and SC coexist, the SC
gap function is the AFM+D-wave gap and  we can fit the data with Born limit
impurities of $\Gamma /\Delta_0=0.32$ (Fig.4 (a)).  At 2.1 GPa where the AFM
disappears, the SC gap becomes the standard harmonic D-wave gap. Assuming that
the Born limit impurity concentration remains the same as in the 1.6 GPa
sample, the effective impurity  scattering rate reduces to $\Gamma
/\Delta_0=0.16$, hence the data fits well again with our calculation (Fig.4
(b)).

The remaining question is the origin of the Born limit scatterers and its
amount. We suggest it could be due to  strain in the lattice induced with
pressure and/or AFM domain walls  in CeRhIn$_5$ and  chemical substitution in
CeCu$_2$(Si$_{0.98}$Ge$_{0.02}$)$_2$. The scattering rate $\Gamma
/\Delta_0=0.32$ in CeRhIn$_5$ at 1.6 GPa appears quite large. But considering
$\Delta_0 \propto$ T$_{sc} \sim$ 1 K, this  impurity scattering rate is
actually quite possible even in the nominally  clean sample. In fact, the
measured specific heat coefficients at low temperatures ($C(T)/T$) both at 1.65
GPa and 2.1 GPa ($N_{1.65}(0)/ N_{normal} \sim 0.4$  and $N_{2.1}(0)/
N_{normal} \sim 0$)  \cite{Fisher} are consistent with our calculations of the
residual  DOS (see the insets of Fig.4 (a) and Fig.4 (b)).

\section{Conclusion}

In summary, we  considered the problem of the  coexistence of AFM and D-wave
SC. Assuming the case of $\Delta_{AFM} > \Delta_{SC}$,  the gap equation is
simplified and the AFM correlation imposes an additional constraint on the
D-wave gap function.  As a result we found an interesting modification of the
D-wave gap function; the D-wave OP should develop additional nodes besides the
original D-wave nodes. The DOS of this AFM+D-wave gap solution has a generic
feature of being almost gapless down to  very low energy. This shape of the DOS
is intrinsically vulnerable to a small amount of impurities to produce  a final
DOS having a large amount of gapless excitations.

We then calculated the nuclear spin-lattice relaxation rate $1/T_1$ with this
AFM+D-wave gap solution  as well as  with a standard harmonic D-wave gap, and
compared the results with the experimental $1/T_1$ of CeRhIn$_5$
\cite{Kitaoka,Mito}. We found that with unitary impurities both the AFM+D-wave
gap and a standard harmonic D-wave gap can produce T-linear behavior in $1/T_1$
at low temperature with a small amount of unitary impurities. However, both
models with unitary impurities cannot explain the 1.6 GPa  and 2.1 GPa data
together in a consistent way.
Then with  Born limit impurities, we could fit the CeRhIn$_5$  $1/T_1$ data at
1.6 GPa  with the AFM+D-wave gap with the amount of impurities $\Gamma
/\Delta_0=0.32$.  Assuming that the gap function is changing from the
AFM+D-wave form at 1.6 GPa to the standard D-wave form at 2.1 GPa, we could
also fit  the  $1/T_1$  data at 2.1 GPa  with the same amount of impurities.
The success of our explanation of the $1/T_1$ nuclear spin-lattice relaxation
rate of CeRhIn$_5$ with pressure supports the idea that a general D-wave gap
function should develop additional nodal points besides the generic D-wave
nodes in the region of coexisting AFM and D-wave SC. These additional nodes
should be observable in carefully designed experiments probing the nodal
quasiparticles \cite{Matsuda}.

\section{Acknowledgement}

We thank  Prof. Y. Onuki, Prof. Kitaoka, Prof. Zheng, Dr. Kawasaki for
providing the experimental data and discussions and  are grateful to Dr. L.H.
Bulaevskii for discussions. This work was supported by US DoE. Y.B. was
partially supported by the Korean Science and Engineering Foundation (KOSEF)
through the Center for Strongly Correlated Materials Research (CSCMR) (2002)
and through the Grant No. 1999-2-114-005-5.

\begin{figure}
\epsfig{figure=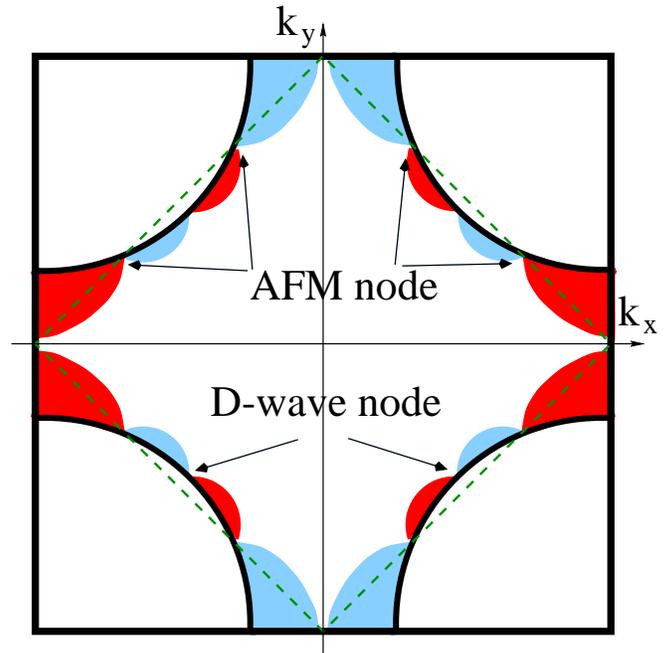,width=1.0\linewidth} \caption{The schematic magnetic
Brillouin zone (BZ) (dashed line) and the Fermi surface (FS) in the
two-dimensional BZ. The crossing points of the magnetic BZ and FS are the AFM
nodal points. Also shown is the schematic gap solution of the AFM+D-wave gap
equation (The red and blue colors indicate the sign-changing OP).
 \label{fig1}}
\end{figure}

\begin{figure}
\epsfig{figure=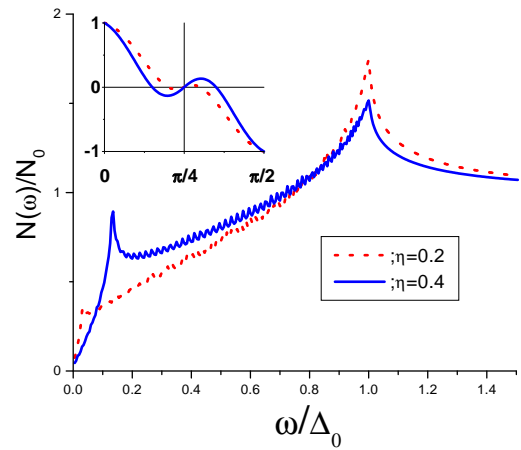,width=1.0\linewidth} \caption{ The normalized DOS
$N(\omega)/N_0$ for different positions of the AFM nodal points $\phi_{AFM}$;
the red dotted line  is for $\eta=|\phi_{AFM}-\phi_{D}|/\phi_{D}$=0.2) and the
blue solid line is for $\eta=0.4$. Inset: The corresponding gap functions
$\Delta(\phi)$.
 \label{fig2}}
\end{figure}

\begin{figure}
\epsfig{figure=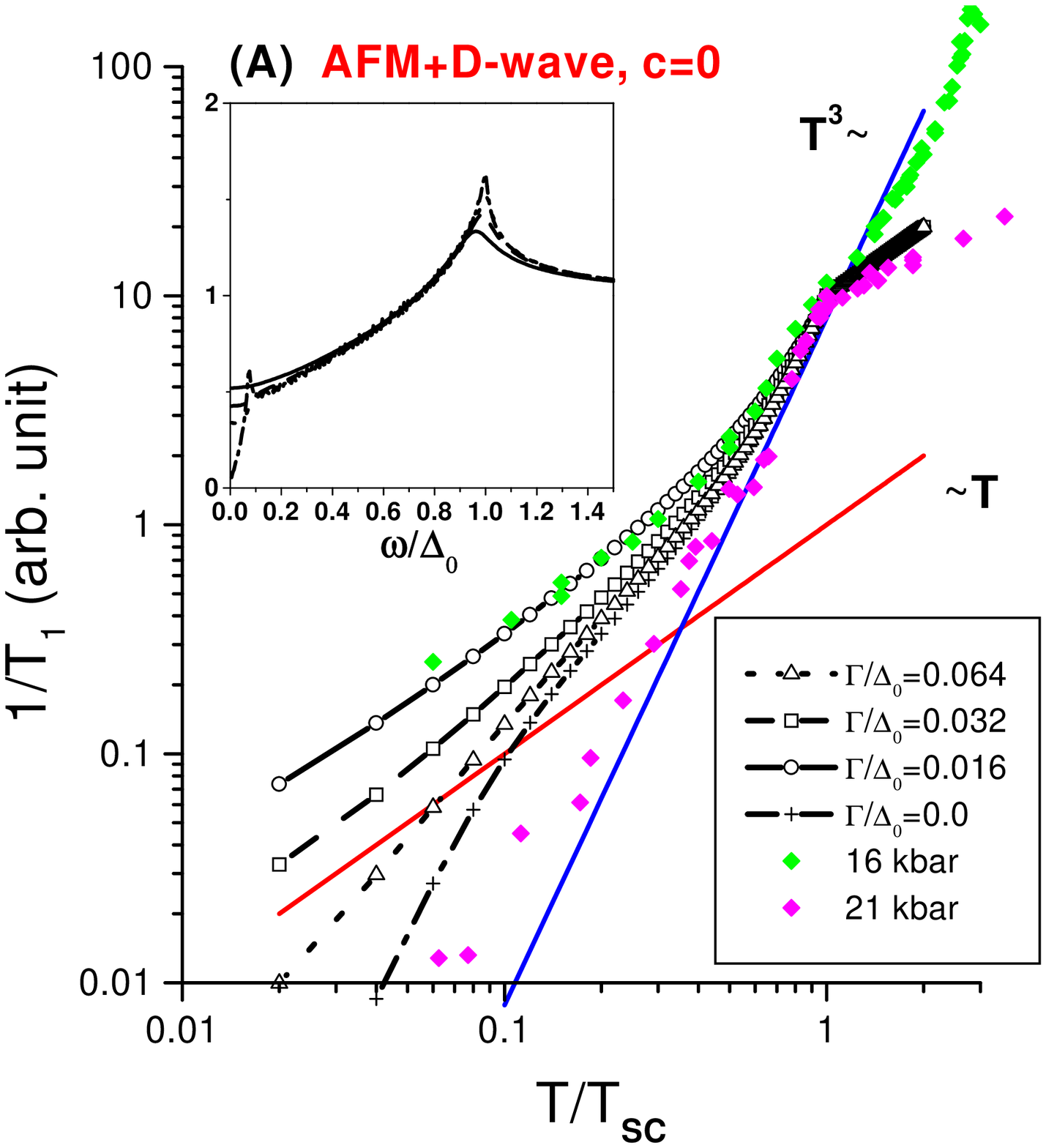,width=1.2\linewidth}
\epsfig{figure=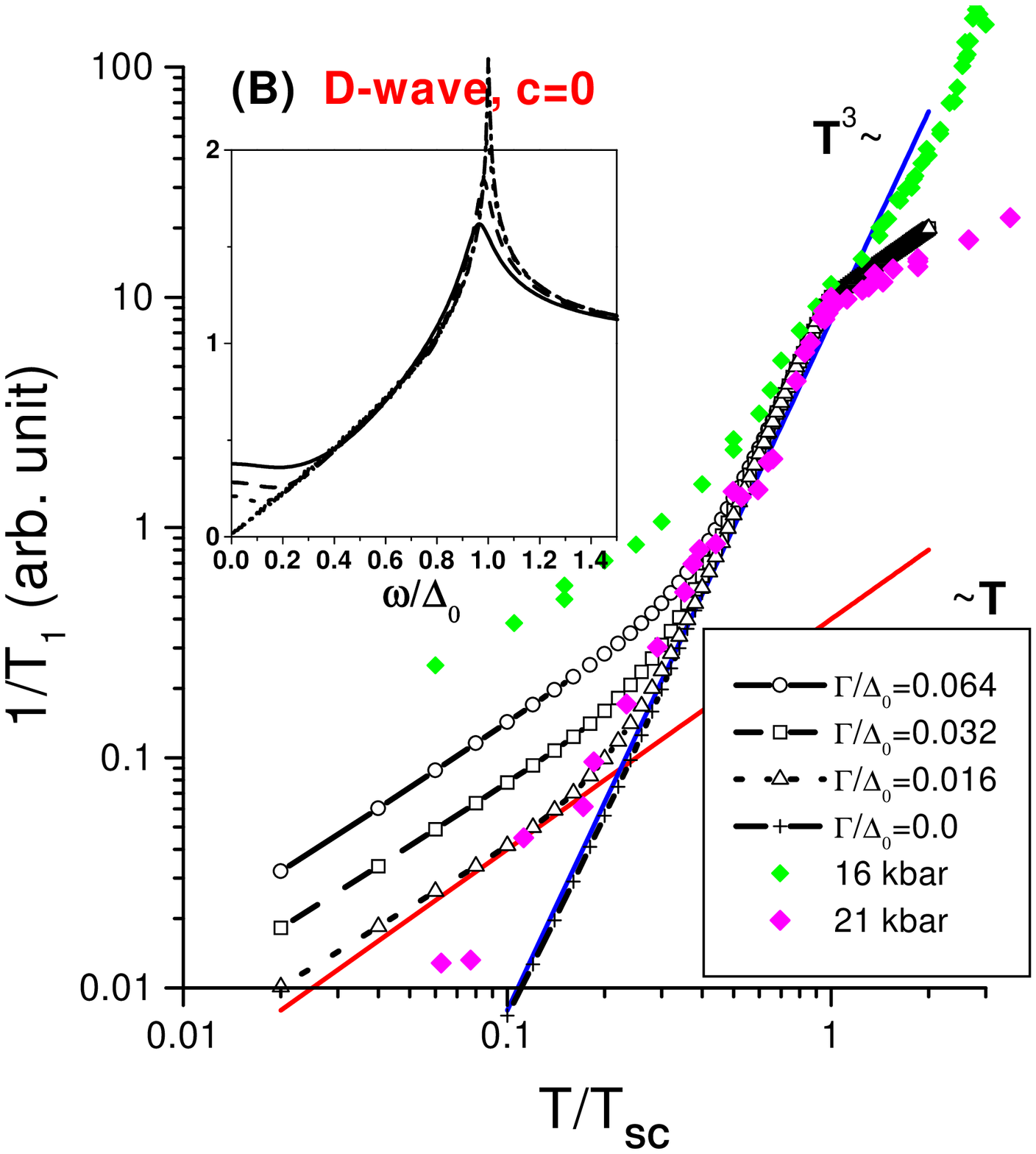,width=1.2\linewidth} \caption{(A) The normalized
1/T$_1$ for AFM+D-wave gap (with the AFM nodal point parameter $\eta$=0.3) for
unitary  scatterer ($c=0$)  with  impurity scattering rates  of
$\Gamma/\Delta_0 =0.064, 0.032, 0.016, 0.0$. The green diamonds are the
normalized 1.6 GPa experimental data and the magenta diamonds are for 2.1 GPa
data of CeRhIn$_5$ \cite{Kitaoka,Mito}. Solid lines for $T^3$ and $T$ are
guides for the eyes. Inset: The corresponding normalized DOS $N(\omega)/N_0$,
in decreasing order of $N(\omega=0)$, with $\Gamma/\Delta_0 =0.064, 0.032,
0.016, 0.0$. (B) The normalized 1/T$_1$ for a standard harmonic D-wave gap for
unitary scatterer ($c=0$)  with impurity scattering rates $\Gamma/\Delta_0
=0.064, 0.032, 0.016, 0.0$. The green and magenta diamonds are the same data as
in Fig.3 (A). Inset: The corresponding normalized DOS $N(\omega)/N_0$.
 \label{fig3}}
\end{figure}

\begin{figure}
\epsfig{figure=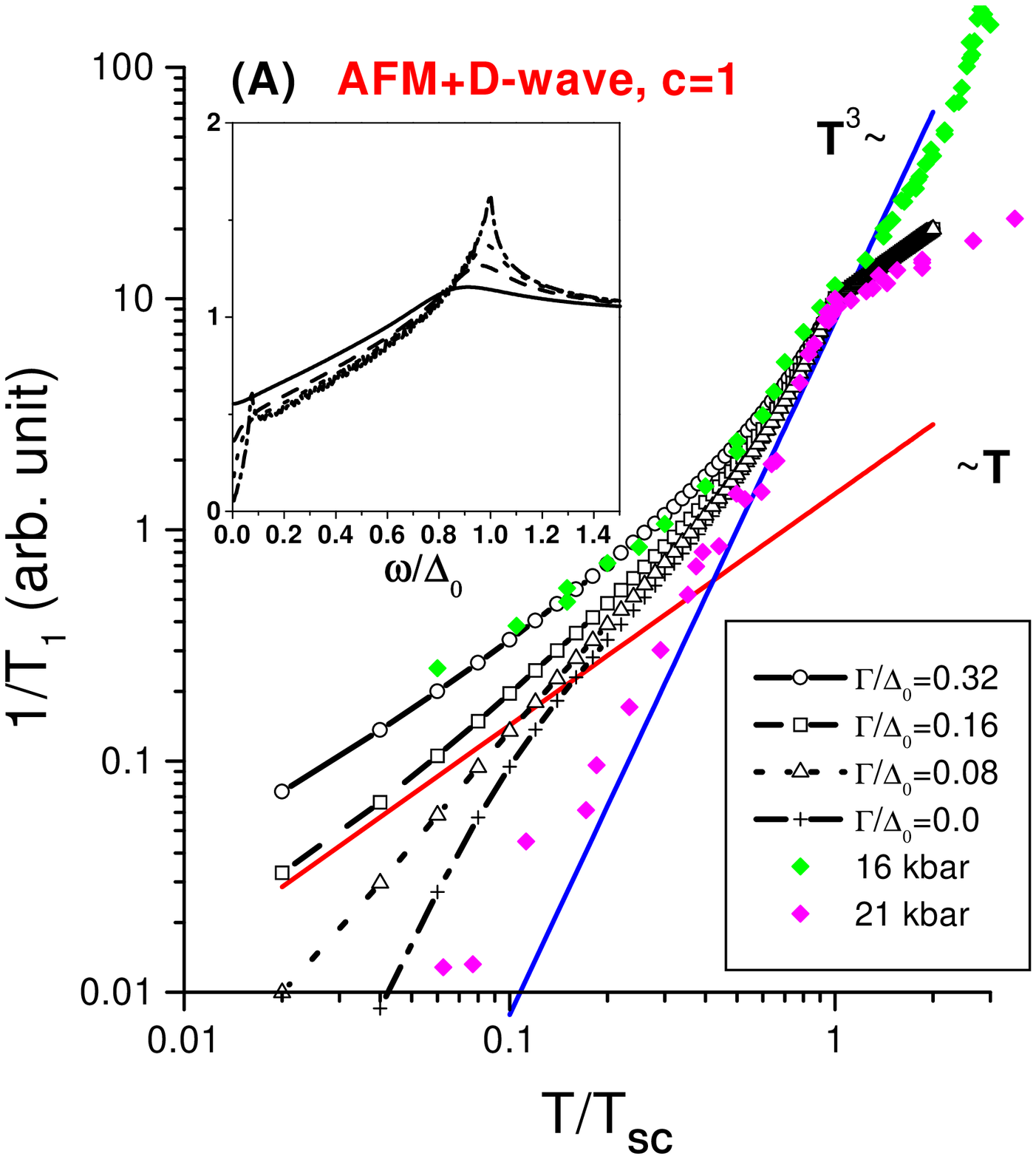,width=1.2\linewidth}
\epsfig{figure=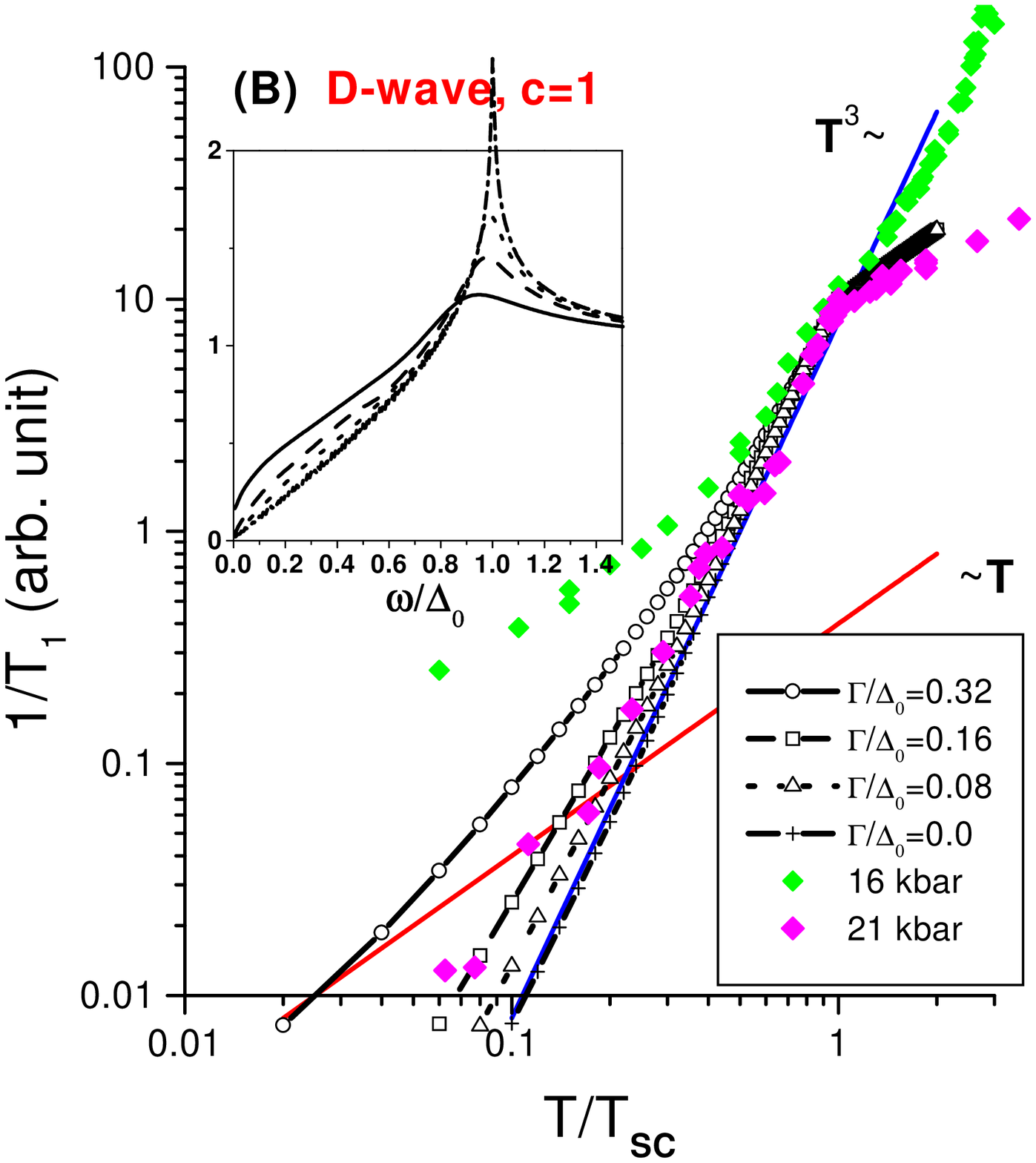,width=1.2\linewidth} \caption{ (A) The normalized
1/T$_1$ for AFM+D-wave gap (with the AFM nodal point parameter $\eta$=0.3) for
Born scatterer ($c=1$)  with impurity scattering rates of $\Gamma/\Delta_0 =
0.32, 0.16, 0.08, 0.0$. The green diamonds are the normalized 1.6 GPa
experimental data and the magenta diamonds are for 2.1 GPa data of CeRhIn$_5$
\cite{Kitaoka,Mito}.  Solid lines  for $T^3$ and $T$ are guides for the eyes.
Inset: The corresponding normalized DOS $N(\omega)/N_0$, in decreasing order of
$N(\omega=0)$, with $\Gamma/\Delta_0 =0.32, 0.16, 0.08, 0.0$. (B) The
normalized 1/T$_1$ for a standard harmonic D-wave gap with Born scatterer
($c=1$)  with impurity scattering rates of $\Gamma/\Delta_0 = 0.32, 0.16, 0.08,
0.0$. The green and magenta diamonds are the same data as in Fig.4 (A). Inset:
The corresponding normalized DOS $N(\omega)/N_0$.
 \label{fig4}}
\end{figure}

\end{multicols}


\begin{references}
\bibitem{pairing potential}
The AFM fluctuations mediated D-wave pairing has been mostly studied in the
paramagnetic phase. In the AFM ordered phase still the longitudinal mode
$\chi_{\parallel}$ can mediate the D-wave pairing and has basically the same
form as in the paramagnetic phase. The issue of pair breaking and pair
mediating by the transverse modes in the AFM phase is beyond the scope of this
work.
\end{references}
\end{document}